\documentclass[onecolumn,trackchanges]{aastex62} 
\usepackage{amsmath,amssymb}

\submitjournal{RAA}

\shorttitle{A brief multi-tracer review}
\shortauthors{Wang \& Zhao}

\def\ie{{\frenchspacing\it i.e.}}
\def\eg{{\frenchspacing\it e.g.}}
\def\etc{{\frenchspacing\it etc.}}

\begin{document}

\title{A brief review on cosmological analysis of galaxy surveys with multiple tracers}

\author{Yuting Wang}\thanks{E-mail: \url{ytwang@nao.cas.cn}}
\affiliation{National Astronomical Observatories, Chinese Academy of Sciences,
             Beijing 100101, P.R.China}
   
\author[0000-0003-4726-6714]{Gong-Bo Zhao}\thanks{E-mail: \url{gbzhao@nao.cas.cn}}
\affiliation{National Astronomical Observatories, Chinese Academy of Sciences,
             Beijing 100101, P.R.China}   
\affiliation{University of Chinese Academy of Sciences, Beijing 100049, P.R.China}

\begin{abstract}
Galaxy redshift surveys are one of the key probes in modern cosmology. In the data analysis of galaxy surveys, the precision of the statistical measurement is primarily limited by the cosmic variance on large scales. Fortunately, this limitation can in principle be evaded by observing multiple types of biased tracers. In this brief review, we present the idea of the multi-tracer method, outline key steps in the data analysis, and show several worked examples based on the GAMA, BOSS and eBOSS galaxy surveys.
\end{abstract}

\section{Introduction}
\label{sec:intro}
Mapping the Universe with massive galaxy surveys provides critical cosmological information, which is key to reveal the physics governing the evolution of the Universe. In particular, galaxy surveys play a crucial role for understanding the origin of the cosmic acceleration discovered in the late 1990s \citep{Riess1998, Perlmutter1999}, which may be due to the existence of Dark Energy (see \citealt{Huterer2018} for a recent review), an energy component with a negative pressure, or to the extension of General Relativity (GR), as reviewed in \cite{Koyama2016}.

Cosmological information in galaxy surveys is primarily encoded in the baryon acoustic oscillations (BAO) and the redshift space distortions (RSD), which are specific clustering patterns of biased tracers \citep{Percival2001, Peacock2001, Eisenstein2005, Cole2005, Percival2007, Beutler2011, Blake2011, Anderson2012, Padmanabhan2012, Ross2015, Alam2017}. Being complementary to other probes including the supernovae (SNIa) \citep{Riess1998, Perlmutter1999, Riess2007, SNLS3, Union, JLA, Pantheon, DESN} and the cosmic microwave background (CMB) \citep{Bennett2003, WMAP9, PLC2018}, the BAO, as a standard ruler, maps the expansion history of the Universe, thus is essential to constrain dark energy, while the RSD, caused by peculiar motions of galaxies, is a natural probe for the nature of gravity on cosmological scales.

\begin{figure*}
\centering
\includegraphics[width=0.8\textwidth]{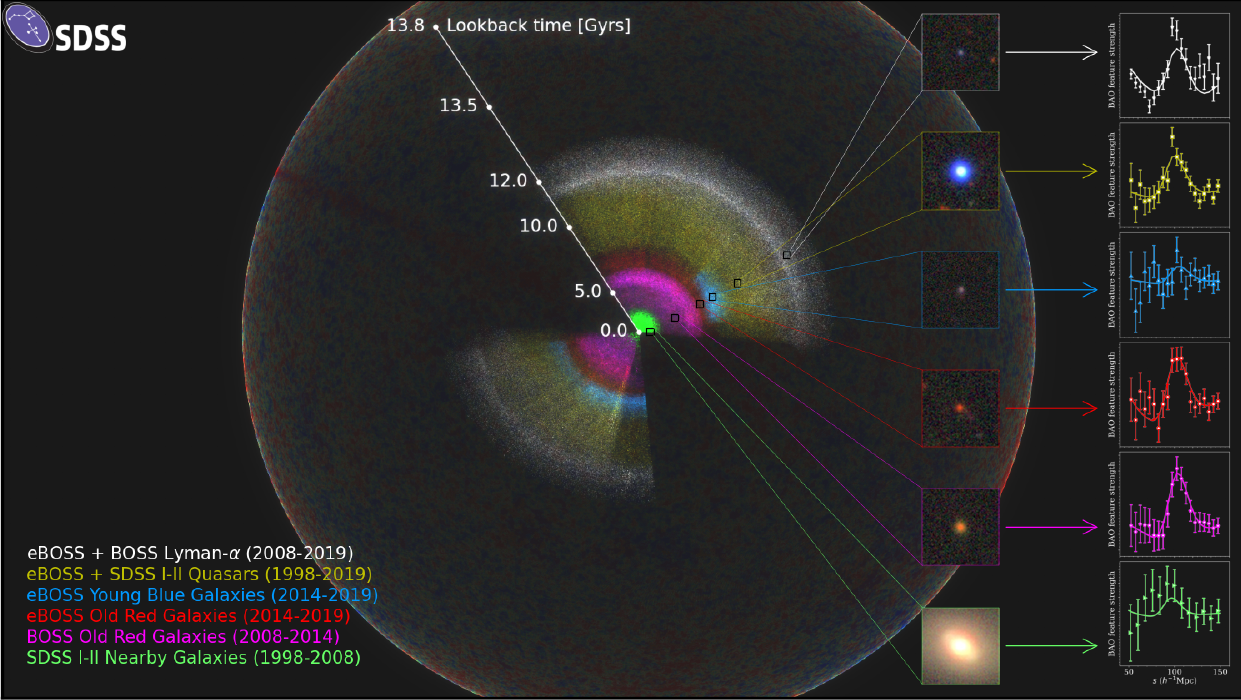}
\caption{Mapping the Universe in three dimensions with SDSS (1998-2020) using different types of tracers including galaxies, quasars, and Lyman-$\alpha$ forest, as illustrated using different colors. The outermost is the last scattering surface of the CMB. The six panels on the right are the BAO-peak measurements using different samples, as shown in legend in the lower-left corner. Image credit: Anand Raichoor (EPFL), Ashley Ross (Ohio State University) and the SDSS Collaboration. This image is a part of the eBOSS DR16 press release, which is available at \url{https://www.sdss.org/press-releases/no-need-to-mind-the-gap/}}
\label{fig:sdssmap}
\end{figure*}

Great progress on galaxy surveys has been made since last century. Started in the 1970's, the Harvard-Smithsonian Center for Astrophysics (CfA) redshift survey brought us the first large galaxy sample, which consists of several thousand galaxies for a clustering analysis \citep{CfAdata,CfA}. Thanks to technical developments of the multi-object spectrographs and charge coupled device (CCD)-based photometry, larger galaxy surveys were built and operated since then, including the Las Campanas Redshift Survey \citep{LCRS}, the Sloan Digital Sky Survey (SDSS) \citep{sdss}, the 2dF Galaxy Redshift Survey (2dFGRS) \citep{2df}, the WiggleZ Dark Energy Survey \citep{wigglez}, the 6dF Galaxy Survey (6dFGS) \citep{6df} and the Galaxy And Mass Assembly (GAMA) survey \citep{GAMA}, \etc

Based on the $2.5$-meter Sloan Telescope \citep{sloantelescope} installed at the Apache Point Observatory, the SDSS project has been one of the most powerful and successful galaxy surveys in the world during the last two decades, with a timeline for her past four generations being: SDSS-I \& II \citep{sdssII} (1998-2008), SDSS-III/Baryon Oscillation Spectroscopic Survey \citep{BOSS} (BOSS; 2008-2014), and SDSS-IV/extended Baryon Oscillation Spectroscopic Survey \citep{eBOSS} (eBOSS; 2014-2020). SDSS has created by far the largest map of the Universe by observing various kinds of tracers, as shown in Fig. \ref{fig:sdssmap}. 

As key quantities for cosmological implications, BAO and RSD are measured from almost all available wide-field galaxy surveys. The statistical uncertainties of BAO and RSD parameters are inherited from the uncertainty in the two-point statistics of galaxy clustering, which consists of two components: the shot noise and the cosmic variance. While the former can be suppressed by increasing the number density of tracers in the observation, the latter cannot be reduced statistically, as long as only one tracer is observed, because of the limited number of pairs on large scales. However, if at least two tracers with an overlapping cosmic volume are observed, the cosmic variance can be reduced to some extent, depending on the level of the shot noise \citep{Seljak2009,McDonald2009}. This `multi-tracer method' works if at least two differently biased tracers of the same underlying dark matter field are available. By comparing the galaxy clustering of these tracers, one is able to determine a ratio between the `effective biases', which may include the RSD parameter $\beta$, and the primordial non-Gaussianity parameter $f_{\rm NL}$, to an infinitive accuracy in the zero noise limit, as detailed in Sec. \ref{sec:method}. The validity of this method with quantitative forecasts has been extensively studied in theory using the Fisher information matrix approach \citep{White2009, Gil-Marin2010, Bernstein2011, Hamaus2011, Abramo2012,Cai2012, Abramo2013, Ferramacho2014, Yamauchi2014, Abramo2016, Zhao2016, Alarcon2018, AA2019,BAA2020,Viljoen2020}. 

Applying this method to actual galaxy surveys requires finding at least two differently biased tracers covering the same patch of the sky and redshift range. This is challenging for most existing surveys, which are designed and optimized for a single tracer within a given footprint and redshift range. One possibility to create such `multi-tracer' samples from a single-tracer survey is to split the sample into subsamples by luminosity or color \citep{Blake2013, Ross2014}, but this may be subject to a limited relative galaxy bias, as samples in a single-tracer survey usually do not differ much in the galaxy bias. Alternatively, one can combine tracers observed by different surveys \citep{Marin2016, Beutler2016}. However, this approach could be limited by the small overlapping area, as most galaxy surveys are designed to be complementary to each other, in terms of the sky coverage and/or redshift range.

Fortunately, the SDSS-IV/eBOSS survey has created a great opportunity for a proper multi-tracer analysis, as it is the first galaxy survey to observe multiple tracers with a large overlap in the cosmic volume. Targeting at both the luminous red galaxies (LRGs, denoted as `Old Red Galaxies' in Fig. \ref{fig:sdssmap} at $0.6<z<1.0$), and emission line galaxies (ELGs, `Young Blue Galaxies' in Fig. \ref{fig:sdssmap} at $0.6<z<1.1$), eBOSS offers observations of the clustering of these two tracers in a large overlapping area (over $700$ deg$^2$) in the same redshift range \citep{eBOSS}. A multi-tracer analysis on the final eBOSS sample, which is tagged as the Data Release (DR)16 data sample, was preformed in both the configuration space \citep{Wang2020} and Fourier space \citep{Zhao2020}.

This article serves as a brief review of the multi-tracer analysis of galaxy surveys, including the methodology (Sec. \ref{sec:method}), key steps in the analysis (Sec. \ref{sec:procedure}), and recent cosmological implications (Sec. \ref{sec:cosmology}). Sec. \ref{sec:conclusion} is devoted to summary and a future outlook.

\section{The multi-tracer method}
\label{sec:method}
We start with a matter over-density field, namely, $\delta_{\rm m} \equiv (\rho- \bar{\rho})/ \bar{\rho} $ with $\bar{\rho}$ being the mean density. The observed galaxy traces the matter field up to an effective bias factor $b_{\rm g}$, relating the over-density of the observed galaxy $\delta_{\rm g}$ to matter $\delta_{\rm m}$ on linear scales, $\ie$, \ $\delta_{\rm g}=b_{\rm g}\delta_{\rm m}$. The measured 2-point statistics of $\delta_{\rm g}$, $\eg$, the power spectrum, $P_g \equiv \langle \delta_{\rm g} \delta_{\rm g} \rangle = b_{\rm g}^2 P_{\rm m}$, is subject to the cosmic variance and the shot noise, which dominates the error budget on the large and small scales, respectively.

If two tracers for the same underlying matter density field are available, then the ratio between $\delta_{\rm g2}$ and $\delta_{\rm g1}$ is,
\begin{eqnarray}
\frac{\delta_{\rm g2}}{\delta_{\rm g1} }= \frac{b_{\rm g2}\delta_{\rm m}+\epsilon_{\rm g2}}{b_{\rm g1}\delta_{\rm m}+\epsilon_{\rm g1}}\,,
\end{eqnarray}
where $\epsilon$ denotes the Poisson noise. We can immediately see that in the low-noise limit, $\ie$, $\epsilon\rightarrow0$, the ratio becomes $\delta_{\rm g2}/\delta_{\rm g1} = b_{\rm g2} /b_{\rm g1}$, thus is free from the cosmic variance, as $\delta_{\rm m}$ is cancelled out. Because the RSD parameter $\beta\equiv f/b$ and the primordial non-Gaussianity parameter $f_{\rm NL}$ are part of the effective bias here, the multi-tracer method can in principle improve the constraint on $\beta$ \citep{McDonald2009} and $f_{\rm NL}$ \citep{Seljak2009}. In practice, the dependence on $\delta_{\rm m}$ can not be completely eliminated due to the shot noise, thus the gain from multiple tracers can be downgraded by various factors including the signal-to-noise ratio of each tracer, the overlapping volume and the relative bias, $\etc$ \ \citep{Gil-Marin2010}.

The general data covariance matrix for a 2-tracer system is, 
\begin{equation} 
C \equiv\left[\begin{array}{cc}
\left\langle\delta_{\rm g 1}^{2}\right\rangle & \left\langle\delta_{\rm g 1} \delta_{\rm g 2}\right\rangle \\
\left\langle\delta_{\rm g 2} \delta_{\rm g 1}\right\rangle & \left\langle\delta_{\rm g 2}^{2}\right\rangle
\end{array}\right]\,,
\end{equation}
and the Fisher matrix can be evaluated as \citep{Tegmark1997}, 
\begin{equation} 
\label{eq:fisher}
F_{\lambda \lambda^{\prime}}=\frac{1}{2} {\rm Tr}\left[C_{, \lambda} C^{-1} C_{, \lambda^{\prime}} C^{-1}\right]
\end{equation}
where $C_{,\lambda} \equiv d C / d \lambda$.

In what follows, we shall apply this result to cases of RSD and the primordial non-Gaussianity, respectively.

\subsection{Determining the RSD parameter with multi-tracer surveys}
In redshift surveys, the radial distance of a galaxy is inferred from the observed redshift, which is determined by the underlying Hubble flow and the peculiar velocity of a galaxy along the line of sight. Hence, the observed galaxy position in redshift space, $x^{\rm s}$ is 
\begin{equation} 
x^{\rm s}=x+\frac{\boldsymbol{v} \cdot \hat{r}}{H},
\end{equation}
where $x$ is the galaxy position in real-space. $\boldsymbol{v} $ the peculiar velocity of a galaxy, and $\hat{r}$ the unit vector along the line of sight. Statistically there is an enhancement of galaxy clustering along the line of sight on large scales due to this peculiar motion, dubbed the RSD. According to the linear perturbation theory, relation between the velocity and matter over-density reads,
\begin{equation} 
-i k^2 \boldsymbol{v} = f H \delta_{\rm m}\,\boldsymbol{k} \,,
\end{equation}
where $f = d \ln \delta_{\rm m} /d \ln a$ is the growth rate of structure, and $\boldsymbol{k}$ is the wave-number in Fourier space. Then the redshift-space galaxy density fluctuation on linear scales is given by \citep{Kaiser},
\begin{eqnarray}
\delta_{\rm g}= \left(b_{\rm g}+f \mu^2\right) \delta_{\rm m}+\epsilon_{\rm g}\,,
\end{eqnarray}
where $\mu$ is the cosine of the angle between $\boldsymbol{k}$ and the line of sight. Now the data covariance matrix becomes,
\begin{equation}
C=\frac{P_{\theta \theta}}{2}\left[\begin{array}{cc}
\left(\beta^{-1}+\mu^{2}\right)^{2} & \left(\beta^{-1}+\mu^{2}\right)\left(\alpha \beta^{-1}+\mu^{2}\right)  \\
\left(\beta^{-1}+\mu^{2}\right)\left(\alpha \beta^{-1}+\mu^{2}\right) & \left(\alpha \beta^{-1}+\mu^{2}\right)^{2}
\end{array}\right]+\frac{N}{2},
\end{equation} where $\beta\equiv f/b_{\rm g1}, \ \alpha\equiv b_{\rm g2}/b_{\rm g1}, \ P_{\theta \theta} \equiv 2 f^{2}\left\langle\delta_{\rm m}^{2}\right\rangle$, and $N_{i j} \equiv 2\left\langle\epsilon_{i} \epsilon_{j}\right\rangle$. Denoting $X_{i j}=N_{i j} / b_{i} b_{j} P_{\rm m}$, \cite{McDonald2009} shows that using only the transverse and the radial modes ($\ie$ \ the modes with $\mu=0$ and $\mu=1$), the uncertainty of $\beta$ is (with $\alpha$ and $P_{\theta\theta}$ marginalized over),

\begin{equation}
\frac{\sigma_{\beta}^{2}}{\beta^{2}}=\frac{\left[\alpha^{2}(1+\beta)^{2}+(\alpha+\beta)^{2}\right] X_{11}-2\left[\alpha^{2}(1+\beta)^{2}+\alpha(1+\beta)(\alpha+\beta)\right] X_{12}+2 \alpha^{2}(1+\beta)^{2} X_{22}}{\beta^{2}(\alpha-1)^{2}},
\end{equation} which goes to zero if $X\rightarrow0$, meaning that $\beta$ can be measured without the cosmic variance from a multi-tracer survey.

\subsection{Determining $f_{\rm NL}$ with multi-tracer surveys}
\begin{figure}
\centering
\includegraphics[width=0.6\textwidth]{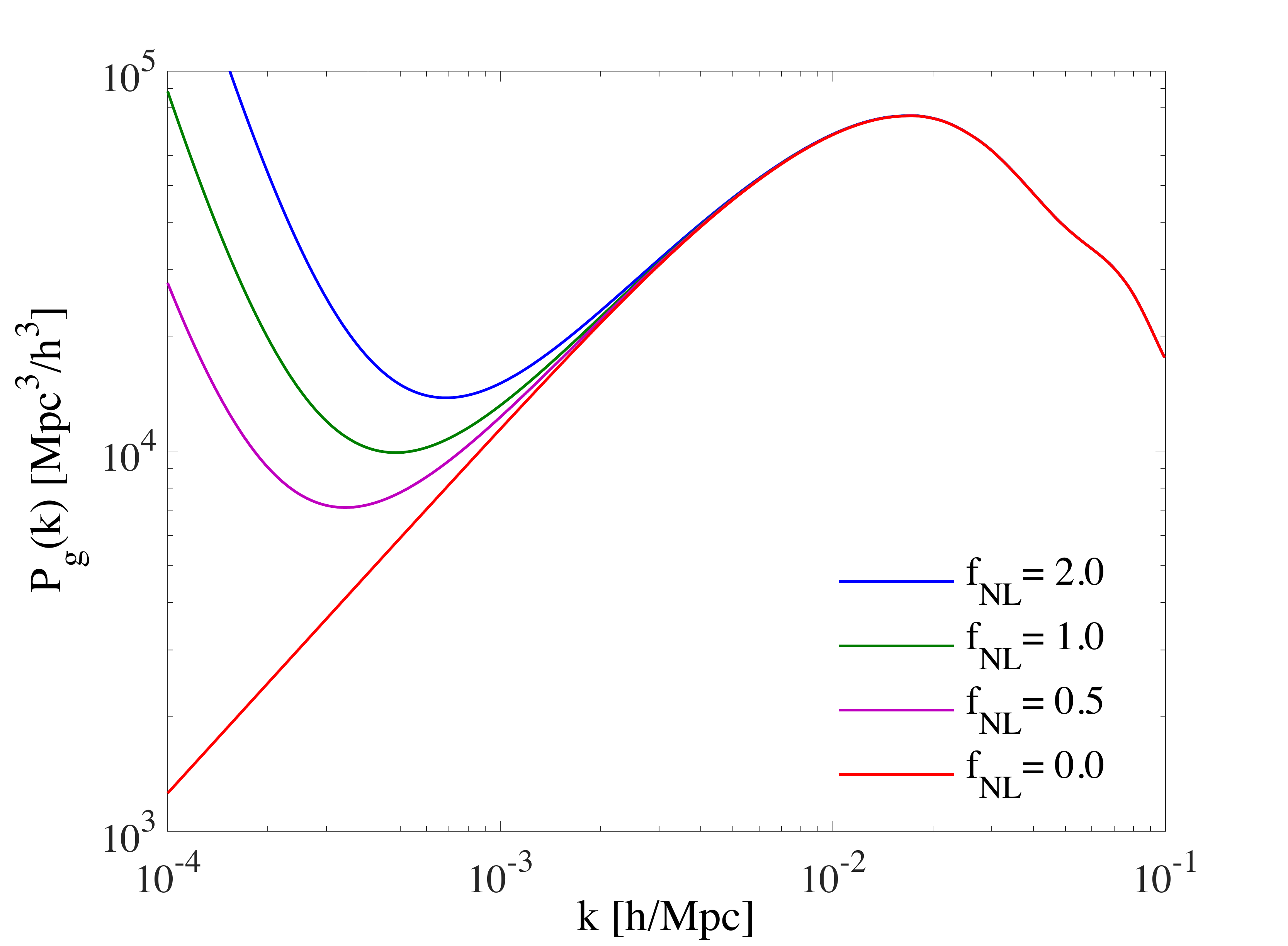}
\caption{The galaxy power spectrum for different values of $f_{\rm NL}$, as shown in the legend.}
\label{fig:pkfnl}
\end{figure}

The local-type of the primordial non-Gaussianity (see \citealt{Wands2010} for a review) can be described by a quadratic correction to the Gaussian field $\phi$, $\ie$, \ $\Phi=\phi+f_{\mathrm{NL}}\left(\phi^{2}-\left\langle\phi^{2}\right\rangle\right)$, in which $f_{\rm NL}$ describes the amplitude of the non-Gaussian correction. This leads to a scale-dependent galaxy bias, $\ie$, \ $b_{\rm g} \rightarrow b_{\rm g}+\Delta b(k)$ with \citep{Dalal2008, Slosar2008}, 
\begin{eqnarray}
\Delta b(k)=3 f_{\mathrm{NL}}(b_{\rm g}-p) \delta_{c} \frac{\Omega_{\rm m}}{k^{2} T(k) D(z)}\left(\frac{H_{0}}{c}\right)^{2}\,,
\end{eqnarray}
where $p$ depends on the type of tracer, $\delta_c$ is the critical linear over-density for the spherical collapse, $T(k)$ is the matter transfer function (normalized to unity on large scales), and $D(z)$ is the growth function (normalized to $a$ in the matter-dominated era). The galaxy over-density thus receives a $k$-dependent correction, say, $\delta_{\rm g}= \left[b_{\rm g}+\Delta b(k, f_{\rm NL})\right] \delta_{\rm m}$, which alters the shape of the power spectrum on large scales, as shown in Fig. \ref{fig:pkfnl}. This makes it possible to constrain $f_{\rm NL}$ using galaxy surveys \citep{Nikoloudakis2013, Ross2013, Karagiannis2014, Mueller2019}. 

As $f_{\rm NL}$ primarily affects the large-scale modes, its constraint can be significantly tightened if the cosmic variance is reduced. For example, the ratio of $\sigma({f_{\rm NL}})$ (with other relavant parameters marginalized over) derived from a two-tracer survey to that from a single-tracer survey using only one mode is \citep{Seljak2009}, 
\begin{equation}
\frac{\sigma\left({f_{\rm NL}}\right)_{\rm 2tr}}{\sigma\left({f_{\rm NL}}\right)_{\rm 1tr}}=\sqrt{2\left[\left({n}_{\rm g2} P_{\rm g2}\right)^{-1}+\left({n}_{\rm g1} P_{\rm g1}\right)^{-1}\right]},
\end{equation} which clearly shows that the gain can be significant in the low-noise limit.

\section{The procedure for a multi-tracer analysis}
\label{sec:procedure}

In this section, we show the key steps in the multi-tracer analysis, including the measurement of the 2-point statistics, modeling and parameter estimation.

\subsection{Measuring the galaxy clustering}
Most of the information in the clustering of galaxies is carried by its two-point correlation function, $\xi(\boldsymbol{s})$, or equivalently, the power spectrum in Fourier space, $P(\boldsymbol{k})$.

\subsubsection{The correlation function}

The 2-point correlation function is measured by counting pairs of galaxies at a given comoving separation, $\boldsymbol{s}$, and an unbiased estimator for two tracers, named as $A$ and $B$, is \citep{LSestimator},
\begin{equation}
\label{eq:crossLS}
\xi_{\rm AB}(s,\mu) = \frac{D_{\rm A}D_{\rm B}-D_{\rm A}R_{\rm B} -D_{\rm B}R_{\rm A}+R_{\rm A}R_{\rm B}}{R_{\rm A}R_{\rm B}}\,,
\end{equation} where $DD$, $DR$, and $RR$ are the normalized number pairs of galaxy-galaxy, galaxy-random and random-random, respectively, within a separation whose central value is $s$, and the cosine between the concerning pair and the line of sight vector is $\mu$. The commonly-used estimator for a single tracer system is a special case of Eq. (\ref{eq:crossLS}) with $A=B$, namely,
\begin{equation}
\label{eq:autoLS}
\xi(s,\mu) = \frac{DD-2DR+RR}{RR}.
\end{equation} 

\begin{figure}
\centering
\includegraphics[width=\textwidth]{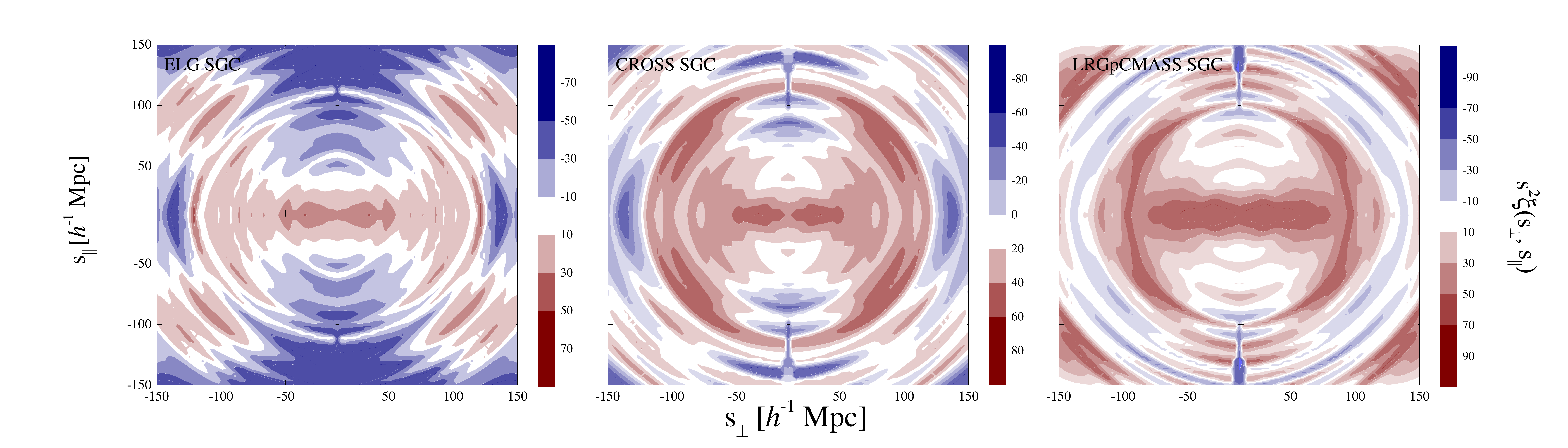}
\caption{The 2D auto- and cross-correlation functions $\xi(s,\mu)$ assembled using the monopole, quadrupole and hexadecapole measured using the estimator in Eq. (\ref{eq:crossLS}), where $s^2= s_{\|}^2+ s_{\perp}^2$, from the samples of eBOSS DR16 ELG (left), LRG (right), and their cross-correlation (middle) in South Galactic Gap. This figure is adopted from \cite{Wang2020}.}
\label{fig:ebossxiSGC}
\end{figure}

Fig. \ref{fig:ebossxiSGC} shows the 2D auto- and cross-correlation functions measured from the eBOSS DR16 LRG and ELG samples using the above estimator Eq. (\ref{eq:crossLS}).

\begin{figure}
\centering
\includegraphics[width=0.9\textwidth]{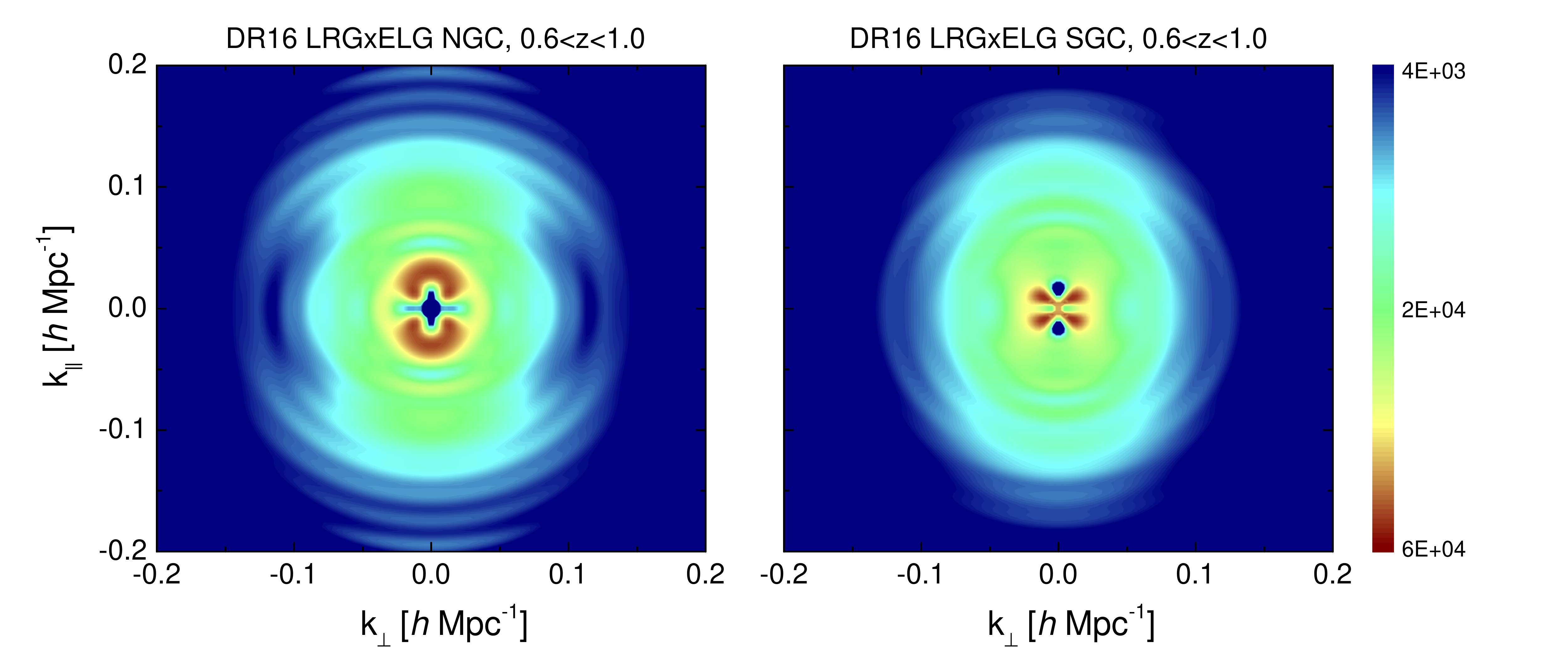}
\caption{The cross power spectrum between the eBOSS DR16 LRGs and ELGs in the Northern (left) and Southern Galactic Cap (right) measured using the estimator Eq. (\ref{eq:PAB}). This figure is adopted from \cite{Zhao2020}.}
\label{fig:Pk2D}
\end{figure}

\subsubsection{The power spectrum}
\label{sec:pkobs}

In Fourier space, the estimated power spectrum with shot noise corrected is given in \citep[FKP estimator;][]{FKPestimator} 
\begin{equation}
\widehat{P}(\boldsymbol{k})=\left\langle|F(\boldsymbol{k})|^{2}\right\rangle-P_{\text {shot }} \,,
\end{equation} 
where $F(\boldsymbol{k})$ is the Fourier transformation of the weighted galaxy fluctuation field, $F(\boldsymbol{r})$ 
\begin{eqnarray}
F(\boldsymbol{r})=\frac{w(\boldsymbol{r})}{I^{1 / 2}}\left[n(\boldsymbol{r})-\alpha n_{\mathrm{s}}(\boldsymbol{r})\right] \,,
\end{eqnarray}
with $n(\boldsymbol{r})$ and $n_{\mathrm{s}}(\boldsymbol{r})$ being the observed number density field for the galaxy catalog and synthetic random catalog, respectively, and $I$ is a normalisation factor. The estimator for the auto- and cross-power spectrum for tracers $A$ and $B$ is \citep{Zhao2020},
\begin{eqnarray} \label{eq:PAB}
\widehat{P}_{\ell}^{\rm AB}(k)=\frac{2 \ell+1}{2 I_{\rm AB}} \int \frac{\mathrm{d} \Omega_{k}}{4 \pi}\left[F_{0, {\rm A}}(\mathbf{k}) F_{\ell, {\rm B}}(-\mathbf{k})+F_{0, {\rm B}}(\mathbf{k}) F_{\ell, {\rm A}}(-\mathbf{k})\right],
\end{eqnarray} with
\begin{eqnarray} \label{eq:FAB}
F_{\ell}(\mathbf{k}) \equiv \int \operatorname{dr} F(\mathbf{r}) e^{i \mathbf{k} \cdot \mathbf{r}} \mathcal{L}_{\ell}(\hat{\mathbf{k}} \cdot \hat{\mathbf{r}}) 
=\frac{4 \pi}{2 \ell+1} \sum_{m=-\ell}^{\ell} Y_{\ell m}(\hat{\mathbf{k}}) \int \operatorname{dr} F(\mathbf{r}) Y_{\ell m}^{*}(\hat{\mathbf{r}}) e^{i \mathbf{k} \cdot \mathbf{r}}.
\end{eqnarray} This is based on the Yamamono estimator \citep{Yamaestimator}, and makes use of the Addition Theorem to reduce the number of Fast Fourier Transform (FFTs) \footnote{The FFT library commonly used is avaible at \url{http://www.fftw.org/}} required in the calculation \citep{Hand2017} \footnote{The FFT-based estimator for the power spectrum multipoles was first developed in \cite{Bianchi2015, Scoccimarro2015} using a decomposition of $\mathcal{L}_{\ell}(\hat{\mathbf{k}} \cdot \hat{\mathbf{r}})$ in Cartesian coordinates, thus it requires a larger number of FFTs than the estimator shown in Eqs. (\ref{eq:PAB}) and (\ref{eq:FAB}).}. The cross power spectrum, measured from the eBOSS DR16 samples using Eq. (\ref{eq:PAB}), is shown in Fig. \ref{fig:Pk2D}.

For the power spectrum analysis, one has to model the effect from the survey geometry carefully, as this window function effect can alter the power spectrum multipoles \footnote{Note that the window function has no effect on the correlation function, because the effect changes the $DD(DR)$ and the $RR$ pairs in the same, multiplicative way, thus it is perfectly cancelled out in the Landy \& Szalay estimator shown in Eq. (\ref{eq:crossLS}).}, as formulated in \cite{Wilson2017}. The survey window function can be measured from the randoms by a pair-counting approach \citep{Wilson2017,Zhao2020,LRG2020}, namely,
\begin{equation}\label{eq:W}
W_{\ell}^{\mathrm{AB}}(s)=\frac{(2 \ell+1)}{I_{\rm AB} \eta^{-2}} \sum_{i, j}^{N_{\mathrm{ran}}} \frac{w_{\mathrm{tot}}^{\mathrm{A}}\left(\mathbf{x}_{i}\right) w_{\mathrm{tot}}^{\mathrm{B}}\left(\mathbf{x}_{j}+\mathbf{s}\right)}{4 \pi s^{3} \Delta(\log s)} \mathcal{L}_{\ell}\left(\hat{\mathbf{x}}_{\mathrm{los}} \cdot \hat{\mathbf{s}}\right),
\end{equation} where $\eta$ is the ratio of the weighted numbers of the data and random. Note that the same normalization factor $I_{\rm AB}$ appears in both Eqs. (\ref{eq:PAB}) and (\ref{eq:W}) , to guarantee that the measured and the theoretical power spectrum are normalized in the same way \citep{RIC}.

\subsection{Modeling the galaxy clustering}
\label{sec:modeling}

This section describes the models commonly used for a multi-tracer survey in both configuration space and Fourier space.

\subsubsection{Modeling the 2-point correlation function}
 
In configuration space, the Gaussian Streaming Model (GSM) \citep{Reid2011, Wang2014} is widely used for modeling the full-shape of the 2-point correlation function,
\begin{eqnarray}\label{eq:xi}
1+\xi(s_{\perp}, s_{\parallel})=\int \frac{\rm d y}{ \sqrt{2\pi \left[\sigma^2_{12}(r, \mu)+\sigma^2_{\rm FoG}\right]}}  \left[1+\xi(r)\right] \times \exp \left\{-\frac{\left[s_{\parallel} - y - \mu v_{12}(r)\right]^2}{2 \left[\sigma^2_{12}(r, \mu)+\sigma^2_{\rm FoG}\right] }\right\},
\label{eq:streaming} 
\end{eqnarray}
where $\smash{s_{||} \equiv s \mu}$ and $\smash{s_{\perp}\equiv s \sqrt{(1-\mu^2)}}$ 
denotes the separation of pairs along and across the LOS, respectively; $\xi(r)$ is the real-space correlation function as a function of the real-space separation $r$; $v_{12}(r)$ is the mean infall velocity of galaxies separated by $r$; and $\sigma_{12}(r, \mu)$ is the pairwise velocity dispersion of galaxies. The parameter $\sigma_{\rm FOG}$ is to account for the Fingers-of-God effect on nonlinear scales. Ingredients including $\xi(r),v_{12}(r),\sigma_{12}(r,\mu)$ are evaluated using the Convolution Lagrangian Perturbation Theory (CLPT) \footnote{Publicly available at \url{https://github.com/wll745881210/CLPT_GSRSD}.} \citep{CLPT,Wang2014}, which requires a linear power spectrum, and two bias parameters $\left\langle F^{\prime}\right\rangle$ and $\left\langle F^{\prime\prime}\right\rangle$. The bias parameters can either be treated as independent parameters in the fitting, or be related using the peak-background split argument \citep{Matsubara2008}.

To generalize Eq. (\ref{eq:xi}) for the multi-tracer case, one only needs to perform the following transformation on the bias parameters, as appeared in the CLPT calculation \citep{CLPT,Wang2014},  
 
\begin{equation}
\begin{aligned}
\left\langle F^{\prime}\right\rangle & \rightarrow \frac{1}{2}\left(\left\langle F_{\rm A}^{\prime}\right\rangle+\left\langle F_{\rm B}^{\prime}\right\rangle\right) \\
\left\langle F^{\prime \prime}\right\rangle & \rightarrow \frac{1}{2}\left(\left\langle F_{\rm A}^{\prime \prime}\right\rangle+\left\langle F_{\rm B}^{\prime \prime}\right\rangle\right) \\
\left\langle F^{\prime}\right\rangle^{2} & \rightarrow\left\langle F_{\rm A}^{\prime}\right\rangle\left\langle F_{\rm B}^{\prime}\right\rangle \\
\left\langle F^{\prime \prime}\right\rangle^{2} & \rightarrow\left\langle F_{\rm A}^{\prime \prime}\right\rangle\left\langle F_{\rm B}^{\prime \prime}\right\rangle \\
\left\langle F^{\prime}\right\rangle\left\langle F^{\prime \prime}\right\rangle & \rightarrow \frac{1}{2}\left(\left\langle F_{\rm A}^{\prime}\right\rangle\left\langle F_{\rm B}^{\prime \prime}\right\rangle+\left\langle F_{\rm A}^{\prime \prime}\right\rangle\left\langle F_{\rm B}^{\prime}\right\rangle\right)
\end{aligned}
\end{equation}

Finally, $\xi(s_{\perp}, s_{\parallel})$ needs to be replaced with $\xi(s_{\perp}^{\prime}, s_{\parallel}^{\prime})$ to account for the Alcock-Paczynski (AP) effect \citep{AP}, which is an anisotropy in the clustering if a wrong cosmology is used to convert the redshifts into distances. The AP effect transforms the coordinates in the following way, \begin{equation}
s_{\perp}^{\prime}=\alpha_{\perp} s_{\perp}, \quad s_{\|}^{\prime}=\alpha_{\|} s_{\|},
\end{equation} with \begin{equation}
\alpha_{\perp}=\frac{D_{M}(z) r_{\rm d}^{\mathrm{fid}}}{D_{M}^{\mathrm{fid}}(z) r_{\rm d}}, \quad \alpha_{\|}=\frac{D_{H}(z) r_{\rm d}^{\mathrm{fid}}}{D_{H}^{\mathrm{fid}}(z) r_{\rm d}}.
\end{equation} where $r_{\rm d}$ denotes the sound horizon at recombination, $D_M (z) \equiv (1 + z)D_A(z)$, and $D_A(z)$ is the angular diameter distance. $D_H(z) = c/H(z)$, $H(z)$ is the Hubble expansion parameter. The superscript ‘fid’ denotes the corresponding values in a fiducial cosmology.
 
\subsubsection{Modeling the power spectrum} 
 
In Fourier space, the TNS model \citep{Taruya2010} is proven to be robust up to quasi-nonlinear scales, thus has been widely used by the BOSS and eBOSS collaboration for data analysis \citep{BOSSpk,DR14QHGM,DR14QGBZ,LRG2020,deMattia2020,Zhao2020}. An extension of the TNS model for a multi-tracer survey is recently developed by \cite{Zhao2020},
\begin{equation}
P_{\mathrm{g}}^{\mathrm{AB}}(k, \mu)= D_{\mathrm{FoG}}(k, \mu)\left[P_{\mathrm{g}, \delta \delta}^{\mathrm{AB}}(k)
+2 f \mu^{2} P_{\mathrm{g}, \delta \theta}^{\mathrm{AB}}(k)+f^{2} \mu^{4} P_{\theta \theta}^{\mathrm{AB}}(k) +A^{\mathrm{AB}}(k, \mu)+B^{\mathrm{AB}}(k, \mu)\right],
\end{equation} where
\begin{eqnarray} 
P_{\mathrm{g}, \delta \delta}^{\mathrm{AB}}(k)&=& b_{1}^{\mathrm{A}} b_{1}^{\mathrm{B}} P_{\mathrm{\delta} \delta}(k)+\left(b_{1}^{\mathrm{A}} b_{2}^{\mathrm{B}}+b_{1}^{\mathrm{B}} b_{2}^{\mathrm{A}}\right) P_{\mathrm{b} 2, \delta}(k) 
+\left(b_{\mathrm{s} 2}^{\mathrm{A}} b_{1}^{\mathrm{B}}+b_{\mathrm{s} 2}^{\mathrm{B}} b_{1}^{\mathrm{A}}\right) P_{\mathrm{bs} 2, \delta}(k) \nonumber\\
&&+\left(b_{\mathrm{s} 2}^{\mathrm{A}} b_{2}^{\mathrm{B}}+b_{\mathrm{s} 2}^{\mathrm{B}} b_{2}^{\mathrm{A}}\right) P_{\mathrm{b} 2 \mathrm{s} 2}(k) 
+\left(b_{3 \mathrm{nl}}^{\mathrm{A}} b_{1}^{\mathrm{B}}+b_{3 \mathrm{nl}}^{\mathrm{B}} b_{1}^{\mathrm{A}}\right) \sigma_{3}^{2}(k) P_{\rm m}^{\mathrm{L}}(k) \nonumber\\
&&+b_{2}^{\mathrm{A}} b_{2}^{\mathrm{B}} P_{\mathrm{b} 22}(k)+b_{\mathrm{s} 2}^{\mathrm{A}} b_{\mathrm{s} 2}^{\mathrm{B}} P_{\mathrm{bs} 22}(k)+N_{\mathrm{AB}} \nonumber\\
P_{\mathrm{g}, \delta \theta}^{\mathrm{AB}}(k)&= &\frac{1}{2}\left[\left(b_{1}^{\mathrm{A}}+b_{1}^{\mathrm{B}}\right), P_{\delta \theta}(k)+\left(b_{2}^{\mathrm{A}}+b_{2}^{\mathrm{B}}\right) P_{\mathrm{b} 2, \theta}(k)
+\left(b_{\mathrm{s} 2}^{\mathrm{A}}+b_{\mathrm{s} 2}^{\mathrm{B}}\right) P_{\mathrm{bs} 2, \theta}(k)\right.,  \nonumber \\
&&\left.+\left(b_{3 \mathrm{nl}}^{\mathrm{A}}+b_{3 \mathrm{nl}}^{\mathrm{B}}\right) \sigma_{3}^{2}(k) P_{\rm m}^{\mathrm{L}}(k)\right] \nonumber\\
P_{\mathrm{g}, \theta \theta}(k)&=&P_{\theta \theta}(k),\nonumber \\
D_{\mathrm{FoG}}(k, \mu)&=&\left\{1+\left[k \mu \sigma_{v}\right]^{2} / 2\right\}^{-2}.
\end{eqnarray} The subscripts $\delta$ and $\theta$ are the over-density and velocity divergence fields, respectively, and $P_{\delta\delta},P_{\delta\theta}$ and $P_{\theta\theta}$ denote the quasi-nonlinear auto- or cross-power spectrum, evaluated using tools such as the regularized perturbation theory (RegPT) \citep{RegPT}. $P_{\rm m}^{\mathrm{L}}(k)$ denotes the linear power spectrum, and terms $b_1$ and $b_2$ stand for the linear bias and the second-order local bias, respectively. The second-order non-local bias $b_{\rm s2}$ and the third-order non-local bias $b_{\rm 3nl}$ can be related to the linear bias via \citep{Chan12},
\begin{equation}
b_{\mathrm{s} 2}=-\frac{4}{7}\left(b_{1}-1\right), \ \ 
b_{3 \mathrm{nl}}=\frac{32}{315}\left(b_{1}-1\right).
\end{equation}
The correction $A, B$ terms for a multi-tracer survey requires a non-trivial extension, and we refer the readers to Appendix A of \cite{Zhao2020} for a full derivation and result. \footnote{A Fortran code for computing these correction terms is available at: \url{http://www2.yukawa.kyoto-u.ac.jp/~atsushi.taruya/cpt_pack.html}}
  
As for the correlation functions, the AP effect distorts the power spectrum by changing $(k,\mu)$ to $(k^{\prime},\mu^{\prime})$ in the following way \citep{Ballinger1996},
\begin{equation}
k^{\prime}=\frac{k}{\alpha_{\perp}}\left[1+\mu^{2}\left(\frac{1}{F^{2}}-1\right)\right]^{1 / 2}; \ \ \ \  \mu^{\prime}=\frac{\mu}{F}\left[1+\mu^{2}\left(\frac{1}{F^{2}}-1\right)\right]^{-1 / 2},
\end{equation} where $F=\alpha_{\|} / \alpha_{\perp}$. The theoretical power spectrum multipoles are,
\begin{equation}
P_{\ell}^{\mathrm{AB}}(k)=\frac{(2 \ell+1)}{2 \alpha_{\perp}^{2} \alpha_{\|}} \int_{-1}^{1} \mathrm{d} \mu P_{\mathrm{g}}^{\mathrm{AB}}\left[k^{\prime}(k, \mu), \mu^{\prime}(\mu)\right] \mathcal{L}_{\ell}(\mu), \end{equation} where $\mathcal{L}_{\ell}$ is the Legendre polynomial of order $\ell$.

As mentioned in Sec. \ref{sec:pkobs}, the measured power spectrum multipoles are the true ones convolved with the survey window function shown in Eq. (\ref{eq:W}), therefore the same convolution needs to be applied to the theoretical prediction, for a fair comparison. An efficient way, which is based on the {\tt FFTLog} algorithm \cite{FFTlog} \footnote{Publicly available at \url{https://jila.colorado.edu/~ajsh/FFTLog/}}, for evaluating this convolution is developed in \cite{Wilson2017}.

\subsection{Parameter estimation}

Given measurements of the 2-point correlation function or power spectrum multipoles from a galaxy sample, the corresponding data covariance matrix measured from the mock catalogs (see \citealt{EZmock} for an example for this procedure) and a theoretical template, one can preform a likelihood analysis to estimate parameters including $\alpha_{||},\alpha_{\bot},f\sigma_8$, as well as $f_{\rm NL}$ using efficient algorithms including the Monte Carlo Markov Chain (MCMC), as implemented in the {\tt CosmoMC} \footnote{Publicly available at \url{https://cosmologist.info/cosmomc/}} \citep{CosmoMC} and {\tt Getdist} \footnote{Publicly available at \url{https://github.com/cmbant/getdist}} \citep{Getdist} packages.

\section{Worked examples}
\label{sec:cosmology}

This section shows a few worked examples of recent multi-tracer analyses, based on the GAMA, BOSS and eBOSS surveys, respectively.

\subsection{A multi-tracer analysis for the GAMA survey}

\begin{figure}
\centering
\includegraphics[width=0.9\textwidth]{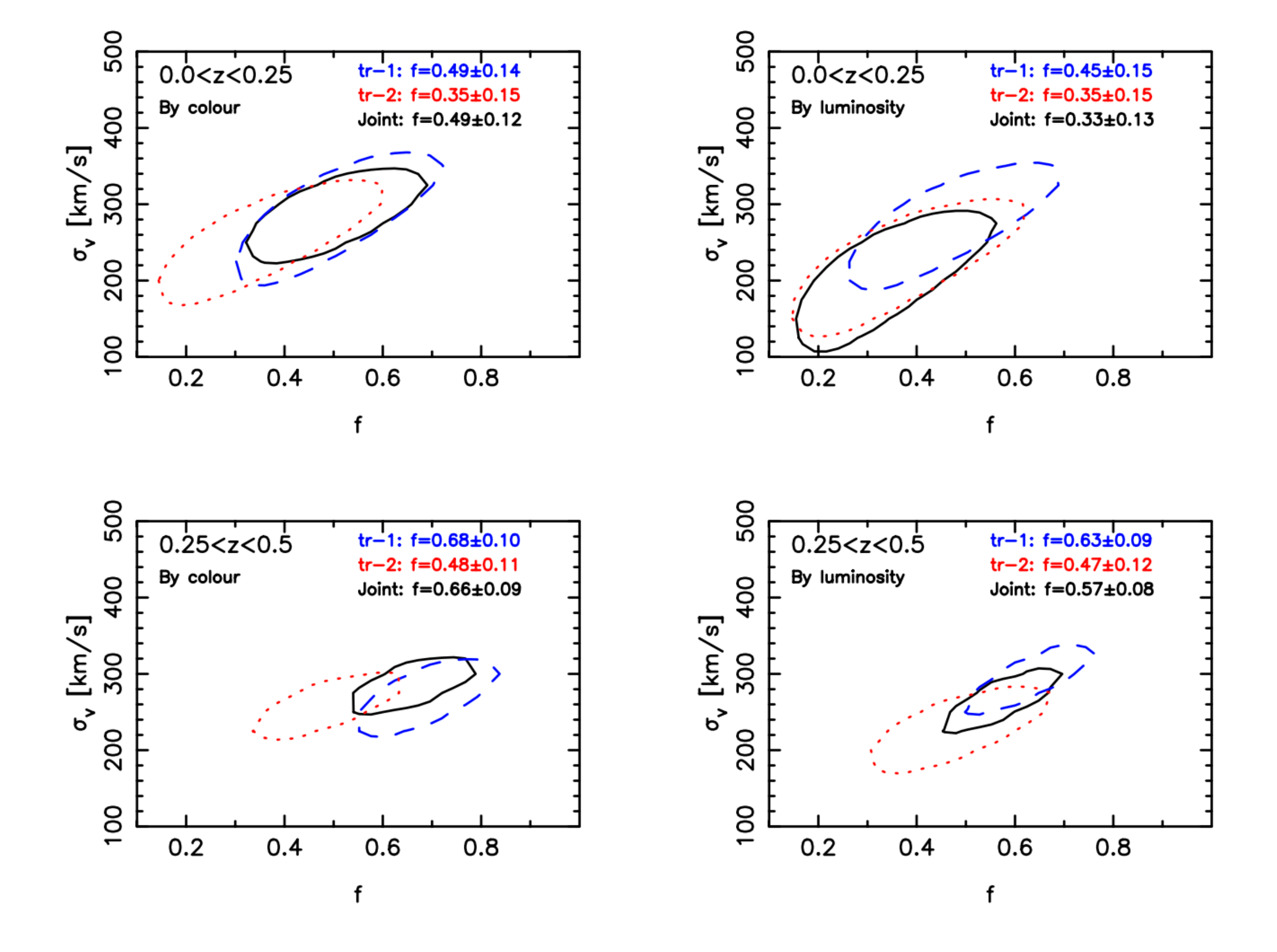}
\caption{The joint constraint on $f$ and $\sigma_{\rm v}$ presented in \cite{Blake2013}. The GAMA sample is split into two using colour (left panels) and luminosity (right), respectively. The blue dashed, red dotted and black solid contours show constraints using tracer 1 (tr-1), tracer 2 (tr-2) and the combined (Joint), respectively. The analysis is performed using galaxies in two redshift slices: $0<z<0.25$ (upper) and $0.25<z<0.5$ (lower).}
\label{fig:gamaf}
\end{figure} 

\begin{figure}
\centering
\includegraphics[width=0.48\textwidth]{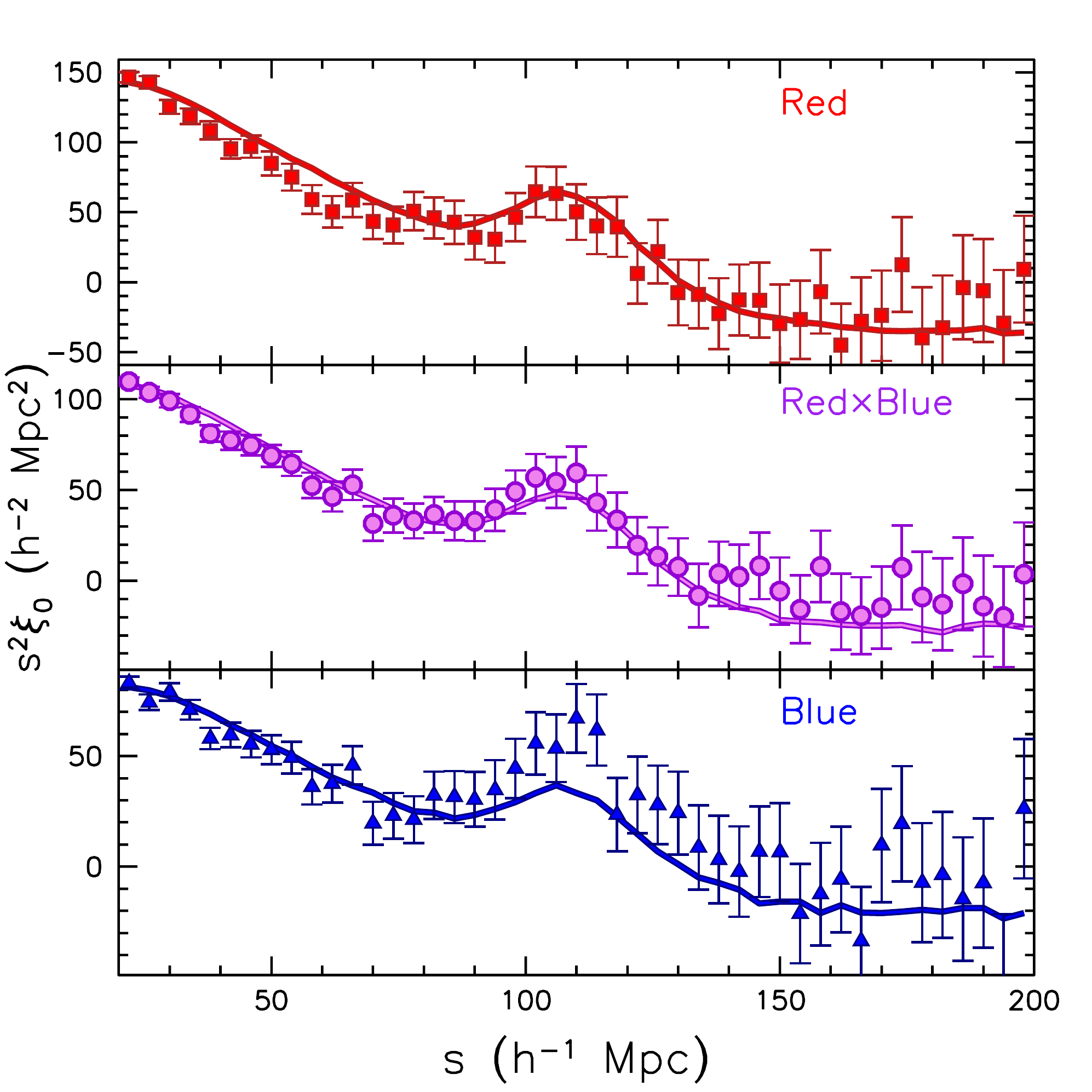}
\includegraphics[width=0.48\textwidth]{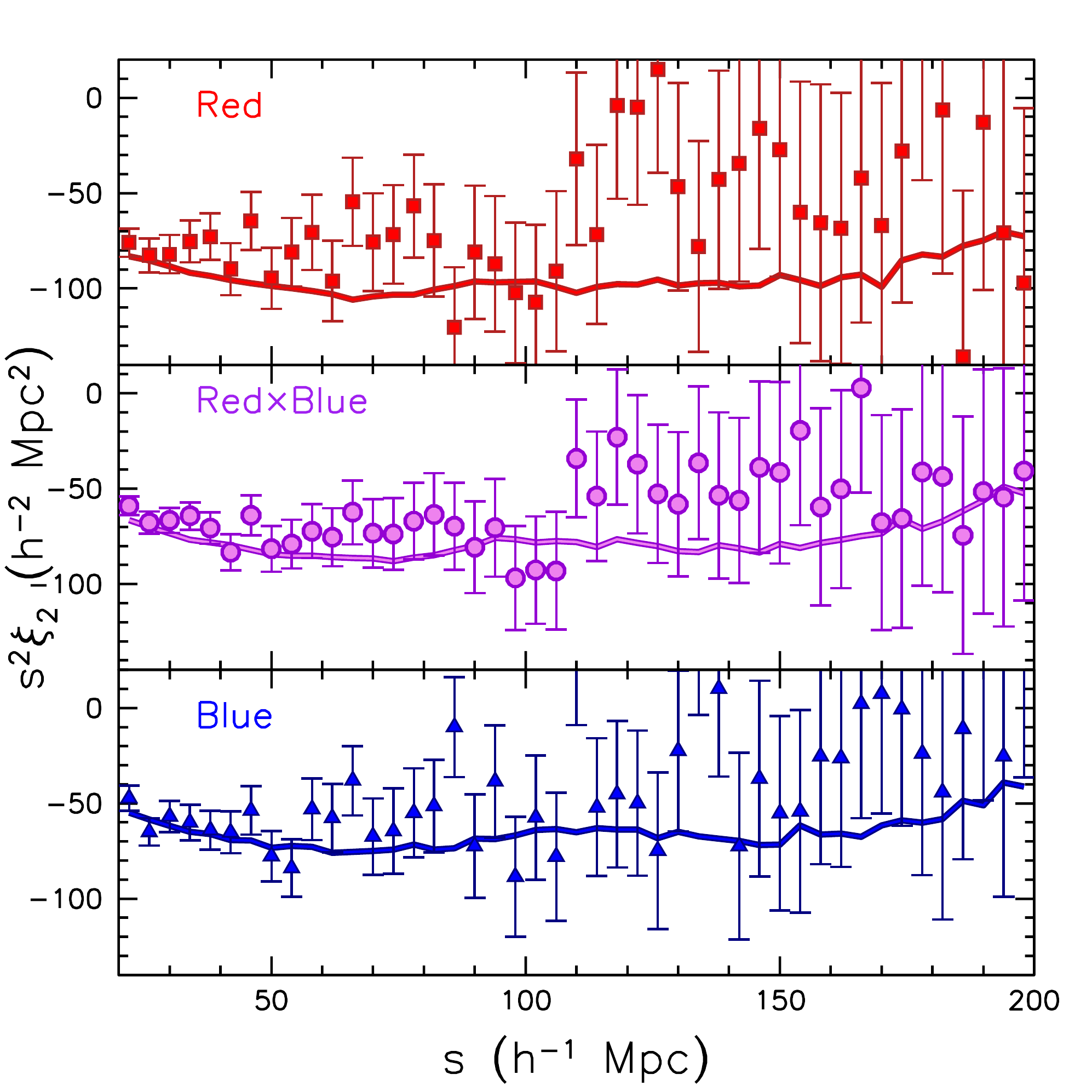}
\caption{The correlation function monopole (left) and quadrupole (right), measured from the BOSS DR10 color-split samples. The top and bottom panels are for the `Red', `Blue' samples, respectively, while the middle panels show the measurement from the cross-correlation. Figure adopted from \citet{Ross2014}.
}
\label{fig:DR10xi}
\end{figure}

The first multi-tracer analysis was performed on the Galaxy and Mass Assembly (GAMA) survey \citep{Blake2013}. GAMA provided a sample of $178,579$ galaxies within the redshift range of $0<z<0.5$ with a coverage area of $180$ deg$^2$. To meet the requirement for a multi-tracer analysis, the entire sample was split into two subsamples by colour and luminosity. A joint constraint on $f$ (the logarithmic growth rate) and $\sigma_{\rm v}$ (the FoG damping parameter) is shown in Fig. \ref{fig:gamaf}, where two subsamples, split by colour (left panels) and luminosity (right panels), respectively, are used for a multi-tracer analysis in two redshift slices $0<z<0.25$ and $0.25<z<0.5$, respectively. It is found that the precision of $f$ from the joint fits (black solid lines) can be improved by $10-20\%$, compared to that from one individual subsample (blue dashed lines for tracer-1 (tr-1) and the red dotted lines for tracer 2 (tr-2). 

\subsection{A multi-tracer analysis for the BOSS survey}

Also splitting samples by the colour, \citet{Ross2014} performed a multi-tracer analysis on the `CMASS' sample released in SDSS-III Baryon Oscillation Spectroscopic Survey (BOSS) DR10. This original CMASS DR10 sample consists of $540, 505$ galaxies within $0.43<z<0.7$. Based on a color-selection, `Red' and `Blue' subsamples are created with a relative bias being $b_{\mathrm{Red}} / b_{\mathrm{Blue}}=1.39 \pm 0.04$. Note that the color-selected samples only have $254,936$ galaxies in total, which halves the number of the original, unsplit sample. 

The monopole and quadrupole of the color-selected samples and their cross-correlation are shown in Fig. \ref{fig:DR10xi}, which yields a constraint on the RSD parameter (with relevant parameters marginalized over) \citep{Ross2014},
\begin{equation}
f \sigma_{8, {\rm Red}}=0.511 \pm 0.083; \ \ \ \ f \sigma_{8, {\rm Blue}}=0.509 \pm 0.085; \ \ \ \ f \sigma_{8, {\rm Cross}}=0.423 \pm 0.061.
\end{equation} This well demonstrates the importance of the cross correlation function for constraining the RSD parameter: it provides a better constraint on $f\sigma_8$ on its own, namely, the uncertainty on $f\sigma_8$ is reduced by $\sim 25\%$ using the cross-correlation function, compared to that using the auto-correlation function. Combining the auto- and cross-correlation further improves the constraint to $f \sigma_{8, \mathrm{comb}}=0.443 \pm 0.055$, which is comparable to that using the original, unsplit sample with doubled number of galaxies, $f \sigma_{8, {\rm full}}=0.422 \pm 0.051$.  

\subsection{A multi-tracer analysis for the eBOSS survey}

As mentioned in Sec. \ref{sec:intro}, the SDSS-IV/eBOSS delivered observations of multiple tracers within one galaxy survey. eBOSS observes three types of discrete tracers, the LRGs, ELGs, and quasars \citep{eBOSS}. The LRGs and ELGs significantly overlap, namely, the overlapping sky coverage is $730\,\rm deg^2$ within the redshift range of $0.6<z<1.0$. Moreover, the LRGs and ELGs have a relative linear bias around $1.5$ \citep{Zhao2020}, making the eBOSS LRG and ELG ideal samples for a multi-tracer analysis.

\begin{figure} 
\centering
\includegraphics[width=0.45\textwidth]{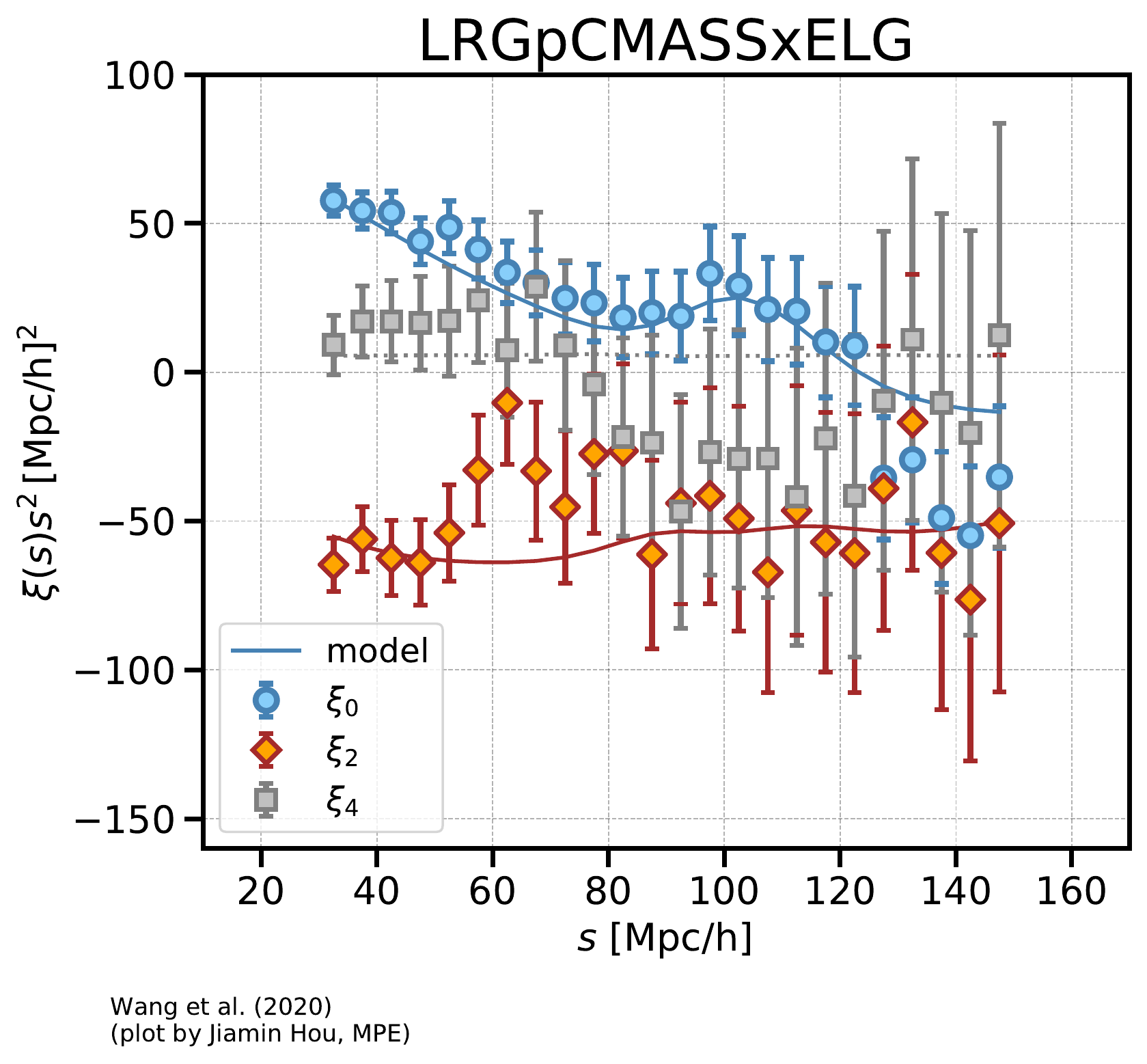} 
\includegraphics[width=0.45\textwidth]{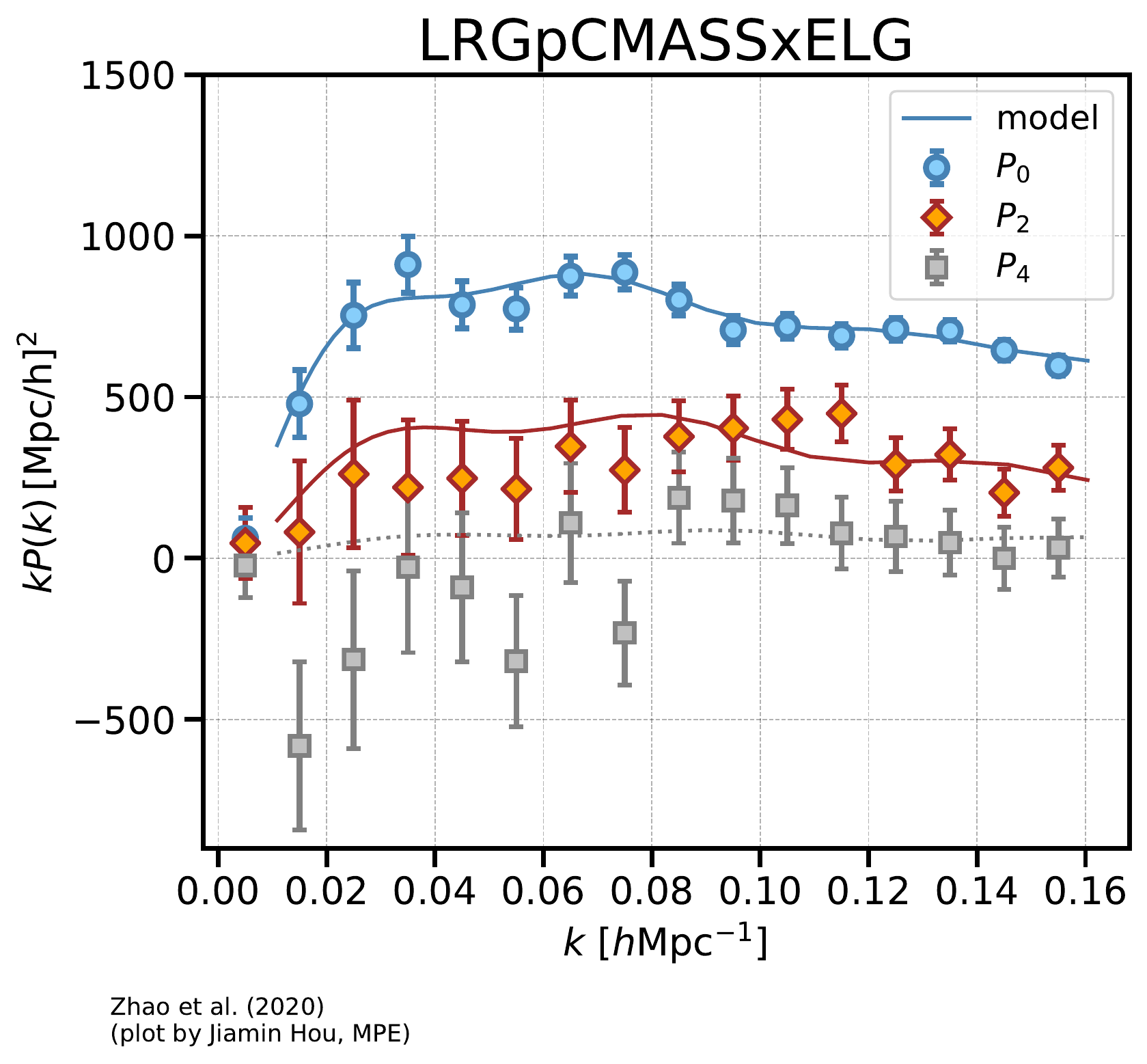} 
\caption{Left: The cross-correlation function multipoles measured from the eBOSS DR16 LRG and ELG samples (data points with error bars), with the best-fit model over-plotted in solid curves \citep{Wang2020}; Right: the corresponding quantities in Fourier space, as analyzed in \cite{Zhao2020}. The plots are made by Jiamin Hou (MPE), and available at \url{https://www.sdss.org/science/final-bao-and-rsd-measurements/}.}
\label{fig:LRGxELG}
\end{figure}

\begin{figure} 
\centering
\includegraphics[width=0.9\textwidth]{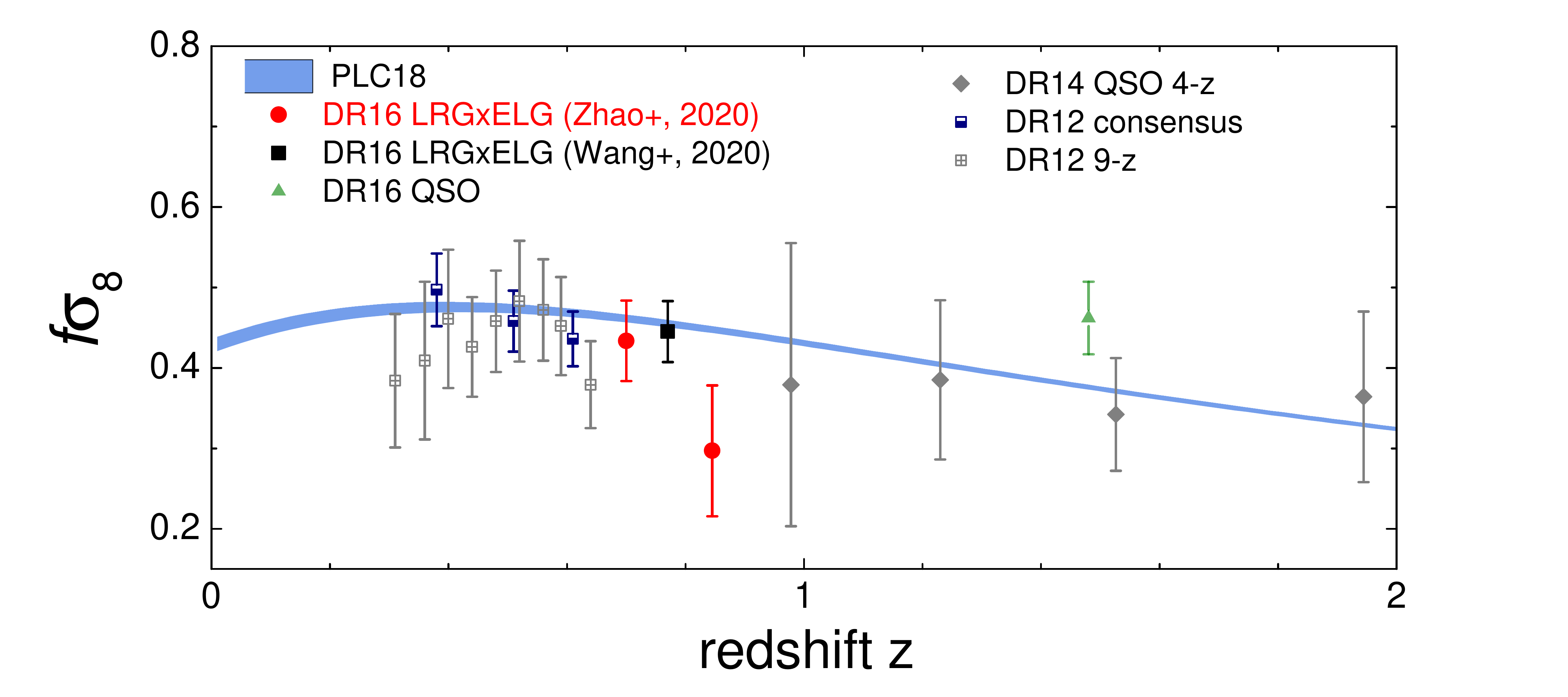}
\caption{A compilation of $f\sigma_8$ measurements from the SDSS BOSS and eBOSS surveys, including those from the latest multi-tracer analysis of the eBOSS DR16 sample in configuration space (DR16 LRG$\times$ELG) \citep{Wang2020} and Fourier-space (DR16 LRG$\times$ELG) \citep{Zhao2020}, the consensus result from eBOSS DR16 QSOs (DR16 QSO) \citep{Hou2020}, the tomographic RSD measurement from the eBOSS DR14 QSO sample using the optimal redshift weighting method (DR14 QSO 4-$z$) \citep{DR14QGBZ}, and the tomographic RSD measurement from the BOSS DR12 sample in Fourier space (DR12 9-$z$) \citep{Zheng2019}. For a reference, the blue band is the 68\% CL constraint derived from Planck 2018 observations in a $\Lambda$CDM cosmology \citep{PLC2018}.}
\label{fig:PLC_rsd}
\end{figure}

A cross-correlation signal is well detected between the eBOSS LRGs and ELGs in both configuration \citep{Wang2020} and Fourier space \citep{Zhao2020}, as shown in Fig. \ref{fig:LRGxELG} (data points with error bars) together with theory curves of the best-fit model (solid lines). The cross-correlation can not only reduce the statistical error of the RSD parameters, it also helps with mitigating the systematic errors, because the contamination in the photometry of each tracer, which is one possible source of the observational systematics, does not correlate with each other. Take the eBOSS ELG sample for example, it is known to be affected by photometric systematics, thus the correlation function and power spectrum of the ELG sample has to be specifically processed for a cosmological analysis \citep{Tamone2020,deMattia2020}. However, the cross correlation function and cross power spectrum are largely immune to this type of systematics, as demonstrated using the mock catalogs \citep{Wang2020,Zhao2020}. 

Fig. \ref{fig:PLC_rsd} shows the measurement of $f\sigma_8$ from the eBOSS multi-tracer analyses, with other recent measurements from BOSS and eBOSS programs. Note that the effective redshifts of the LRG and ELG samples are different, and \cite{Zhao2020} and \cite{Wang2020} take different approaches to account for this effect (both approaches are validated by the mock tests), thus the resultant measurements are at different effective redshifts, $\eg$, \cite{Zhao2020} reports two measurements at $z=0.70$ and $0.845$, respectively, while \cite{Wang2020} only measures at $z=0.77$. While a direct comparison is not straightforward at this level, \cite{Zhao2020} performed an additional measurement at $z=0.77$, and found a consistency with \cite{Wang2020}. In this analysis, it is found that the constraint on $f\sigma_8$ gets improved by approximately $12\%$, compared to that using the LRGs alone \citep{Wang2020}.

\section{Conclusion and discussions}
\label{sec:conclusion}

Stepping into the stage-IV era of galaxy surveys, we are seeking statistical and theoretical methods to improve the precision on the measured cosmological parameters, which is limited by the cosmic variance and the shot noise. One promising way to beat the cosmic variance is to contrast the 2-point statistics of different tracers, yielding a quantity of cosmological importance, which is free from the cosmic variance.

In this brief review, we present the basic idea of the multi-tracer method, outline a procedure for performing the multi-tracer analysis, followed by a few worked examples based on large galaxy surveys, which has well demonstrated the efficacy of this method. 

The newly-started Dark Energy Spectroscopic Instrument \citep[DESI;][]{DESIpaperI, DESIpaperII} is a typical multi-tracer survey, whose LRGs and ELGs targets do overlap significantly across a wide redshift range with a much higher number density than that of eBOSS. This makes it ideal for a multi-tracer analysis for improving the statistical precision of the RSD and primordial non-Gaussianity. The multi-tracer analysis also helps with the observational systematics through the cross-correlation between tracers, in addition to mitigating the systematics by a better modeling and pipeline for processing the raw observations.

\acknowledgments

GBZ is supported by the National Key Basic Research and Development Program of China (No. 2018YFA0404503), and a grant of CAS Interdisciplinary Innovation Team. YW is supported by the Nebula Talents Program of NAOC. YW and GBZ are supported by NSFC Grants 11925303, 11720101004, 11673025 and 11890691.

Funding for the Sloan Digital Sky Survey IV has been provided by the Alfred P. Sloan Foundation, the U.S. Department of Energy Office of Science, and the Participating Institutions. SDSS-IV acknowledges
support and resources from the Center for High-Performance Computing at
the University of Utah. The SDSS web site is \url{http://www.sdss.org/}. 

SDSS-IV is managed by the Astrophysical Research Consortium for the 
Participating Institutions of the SDSS Collaboration including the 
Brazilian Participation Group, the Carnegie Institution for Science, 
Carnegie Mellon University, the Chilean Participation Group, the French Participation Group, Harvard-Smithsonian Center for Astrophysics, 
Instituto de Astrof\'isica de Canarias, The Johns Hopkins University, Kavli Institute for the Physics and Mathematics of the Universe (IPMU) / 
University of Tokyo, the Korean Participation Group, Lawrence Berkeley National Laboratory, 
Leibniz Institut f\"ur Astrophysik Potsdam (AIP),  
Max-Planck-Institut f\"ur Astronomie (MPIA Heidelberg), 
Max-Planck-Institut f\"ur Astrophysik (MPA Garching), 
Max-Planck-Institut f\"ur Extraterrestrische Physik (MPE), 
National Astronomical Observatories of China, New Mexico State University, 
New York University, University of Notre Dame, 
Observat\'ario Nacional / MCTI, The Ohio State University, 
Pennsylvania State University, Shanghai Astronomical Observatory, 
United Kingdom Participation Group,
Universidad Nacional Aut\'onoma de M\'exico, University of Arizona, 
University of Colorado Boulder, University of Oxford, University of Portsmouth, 
University of Utah, University of Virginia, University of Washington, University of Wisconsin, 
Vanderbilt University, and Yale University.


\begin{thebibliography}{}
\expandafter\ifx\csname natexlab\endcsname\relax\def\natexlab#1{#1}\fi
\providecommand{\url}[1]{\href{#1}{#1}}
\providecommand{\dodoi}[1]{doi:~\href{http://doi.org/#1}{\nolinkurl{#1}}}
\providecommand{\doeprint}[1]{\href{http://ascl.net/#1}{\nolinkurl{http://ascl.net/#1}}}
\providecommand{\doarXiv}[1]{\href{https://arxiv.org/abs/#1}{\nolinkurl{https://arxiv.org/abs/#1}}}

\end{thebibliography}


\begin{thebibliography}{99}

\bibitem[Abramo(2012)]{Abramo2012} Abramo, L.~R.\ 2012, \mnras, 420, 2042

\bibitem[Abramo \& Amendola(2019)]{AA2019} Abramo, L.~R. \& Amendola, L.\ 2019, \jcap, 2019, 030

\bibitem[Abramo \& Leonard(2013)]{Abramo2013} Abramo, L.~R. \& Leonard, K.~E.\ 2013, \mnras, 432, 318

\bibitem[Abramo et al.(2016)]{Abramo2016} Abramo, L.~R., Secco, L.~F., \& Loureiro, A.\ 2016, \mnras, 455, 3871

\bibitem[Abazajian et al.(2009)]{sdssII} Abazajian, K.~N., Adelman-McCarthy, J.~K., Ag{\"u}eros, M.~A., et al.\ 2009, \apjs, 182, 543

\bibitem[Alam et al.(2017)]{Alam2017} Alam, S., Ata, M., Bailey, S., et al.\ 2017, \mnras, 470, 2617

\bibitem[Alarcon et al.(2018)]{Alarcon2018} Alarcon, A., Eriksen, M., \& Gaztanaga, E.\ 2018, \mnras, 473, 1444

\bibitem[Alcock \& Paczynski(1979)]{AP} Alcock, C. \& Paczynski, B.\ 1979, \nat, 281, 358


\bibitem[Anderson et al.(2012)]{Anderson2012} Anderson, L., Aubourg, E., Bailey, S., et al.\ 2012, \mnras, 427, 3435


\bibitem[Baldry et al.(2010)]{GAMA} Baldry, I.~K., Robotham, A.~S.~G., Hill, D.~T., et al.\ 2010, \mnras, 404, 86

\bibitem[Ballinger et al.(1996)]{Ballinger1996} Ballinger, W.~E., Peacock, J.~A., \& Heavens, A.~F.\ 1996, \mnras, 282, 877

\bibitem[Bennett et al.(2003)]{Bennett2003} Bennett, C.~L., Halpern, M., Hinshaw, G., et al.\ 2003, \apjs, 148, 1

\bibitem[Bernstein \& Cai(2011)]{Bernstein2011} Bernstein, G.~M. \& Cai, Y.-C.\ 2011, \mnras, 416, 3009

\bibitem[Betoule et al.(2014)]{JLA} Betoule, M., Kessler, R., Guy, J., et al.\ 2014, \aap, 568, A22

\bibitem[Beutler et al.(2011)]{Beutler2011} Beutler, F., Blake, C., Colless, M., et al.\ 2011, \mnras, 416, 3017

\bibitem[Beutler et al.(2016)]{Beutler2016} Beutler, F., Blake, C., Koda, J., et al.\ 2016, \mnras, 455, 3230

\bibitem[Beutler et al.(2017)]{BOSSpk} Beutler, F., Seo, H.-J., Saito, S., et al.\ 2017, \mnras, 466, 2242

\bibitem[Blake et al.(2008)]{wigglez} Blake, C., Brough, S., Couch, W., et al.\ 2008, Astronomy and Geophysics, 49, 5.19

\bibitem[Blake et al.(2011)]{Blake2011} Blake, C., Kazin, E.~A., Beutler, F., et al.\ 2011, \mnras, 418, 1707

\bibitem[Blake et al.(2013)]{Blake2013} Blake, C., Baldry, I.~K., Bland-Hawthorn, J., et al.\ 2013, \mnras, 436, 3089

\bibitem[Bianchi et al.(2015)]{Bianchi2015} Bianchi, D., Gil-Mar{\'\i}n, H., Ruggeri, R., et al.\ 2015, \mnras, 453, L11

\bibitem[Boschetti et al.(2020)]{BAA2020} Boschetti, R., Abramo, L.~R., \& Amendola, L.\ 2020, arXiv:2005.02465

\bibitem[Cai \& Bernstein(2012)]{Cai2012} Cai, Y.-C. \& Bernstein, G.\ 2012, \mnras, 422, 1045


\bibitem[Carlson et al.(2013)]{CLPT} Carlson, J., Reid, B., \& White, M.\ 2013, \mnras, 429, 1674

\bibitem[Chan et al.(2012)]{Chan12} Chan, K.~C., Scoccimarro, R., \& Sheth, R.~K.\ 2012, \prd, 85, 083509

\bibitem[Cole et al.(2005)]{Cole2005} Cole, S., Percival, W.~J., Peacock, J.~A., et al.\ 2005, \mnras, 362, 505

\bibitem[Colless et al.(2001)]{2df} Colless, M., Dalton, G., Maddox, S., et al.\ 2001, \mnras, 328, 1039

\bibitem[Dalal et al.(2008)]{Dalal2008} Dalal, N., Dor{\'e}, O., Huterer, D., et al.\ 2008, \prd, 77, 123514

\bibitem[D'Andrea et al.(2018)]{DESN} D'Andrea, C.~B., Smith, M., Sullivan, M., et al.\ 2018, arXiv:1811.09565

\bibitem[Davis \& Peebles(1983)]{CfA} Davis, M. \& Peebles, P.~J.~E.\ 1983, \apj, 267, 465

\bibitem[Dawson et al.(2016)]{eBOSS} Dawson, K.~S., Kneib, J.-P., Percival, W.~J., et al.\ 2016, \aj, 151, 44

\bibitem[de Mattia \& Ruhlmann-Kleider(2019)]{RIC} de Mattia, A. \& Ruhlmann-Kleider, V.\ 2019, \jcap, 2019, 036

\bibitem[de Mattia et al.(2020)]{deMattia2020} de Mattia, A., Ruhlmann-Kleider, V., Raichoor, A., et al.\ 2020, arXiv:2007.09008

\bibitem[DESI Collaboration et al.(2016 a)]{DESIpaperI} DESI Collaboration, Aghamousa, A., Aguilar, J., et al.\ 2016, arXiv:1611.00036

\bibitem[DESI Collaboration et al.(2016 b)]{DESIpaperII} DESI Collaboration, Aghamousa, A., Aguilar, J., et al.\ 2016, arXiv:1611.00037

\bibitem[Eisenstein et al.(2005)]{Eisenstein2005} Eisenstein, D.~J., Zehavi, I., Hogg, D.~W., et al.\ 2005, \apj, 633, 560

\bibitem[Eisenstein et al.(2011)]{BOSS} Eisenstein, D.~J., Weinberg, D.~H., Agol, E., et al.\ 2011, \aj, 142, 72

\bibitem[Feldman et al.(1994)]{FKPestimator} Feldman, H.~A., Kaiser, N., \& Peacock, J.~A.\ 1994, \apj, 426, 23

\bibitem[Ferramacho et al.(2014)]{Ferramacho2014} Ferramacho, L.~D., Santos, M.~G., Jarvis, M.~J., et al.\ 2014, \mnras, 442, 2511

\bibitem[Gil-Mar{\'\i}n et al.(2020)]{LRG2020} Gil-Mar{\'\i}n, H., Bautista, J.~E., Paviot, R., et al.\ 2020, arXiv:2007.08994

\bibitem[Gil-Mar{\'\i}n et al.(2018)]{DR14QHGM} Gil-Mar{\'\i}n, H., Guy, J., Zarrouk, P., et al.\ 2018, \mnras, 477, 1604

\bibitem[Gil-Mar{\'\i}n et al.(2010)]{Gil-Marin2010} Gil-Mar{\'\i}n, H., Wagner, C., Verde, L., et al.\ 2010, \mnras, 407, 772

\bibitem[Gunn et al.(2006)]{sloantelescope} Gunn, J.~E., Siegmund, W.~A., Mannery, E.~J., et al.\ 2006, \aj, 131, 2332

\bibitem[Hamaus et al.(2011)]{Hamaus2011} Hamaus, N., Seljak, U., \& Desjacques, V.\ 2011, \prd, 84, 083509

\bibitem[Hamilton(2000)]{FFTlog} Hamilton, A.~J.~S.\ 2000, \mnras, 312, 257

\bibitem[Hand et al.(2017)]{Hand2017} Hand, N., Li, Y., Slepian, Z., et al.\ 2017, \jcap, 2017, 002

\bibitem[Hinshaw et al.(2013)]{WMAP9} Hinshaw, G., Larson, D., Komatsu, E., et al.\ 2013, \apjs, 208, 19

\bibitem[Hou et al.(2020)]{Hou2020} Hou, J., S{\'a}nchez, A.~G., Ross, A.~J., et al.\ 2020, arXiv:2007.08998

\bibitem[Huchra et al.(1983)]{CfAdata} Huchra, J., Davis, M., \& Latham, D.\ 1983, Cambridge: Smithsonian Center for Astrophysics, 1983

\bibitem[Huterer \& Shafer(2018)]{Huterer2018} Huterer, D. \& Shafer, D.~L.\ 2018, Reports on Progress in Physics, 81, 016901

\bibitem[Jones et al.(2009)]{6df} Jones, D.~H., Read, M.~A., Saunders, W., et al.\ 2009, \mnras, 399, 683

\bibitem[Kaiser(1987)]{Kaiser} Kaiser, N.\ 1987, \mnras, 227, 1

\bibitem[Karagiannis et al.(2014)]{Karagiannis2014} Karagiannis, D., Shanks, T., \& Ross, N.~P.\ 2014, \mnras, 441, 486

\bibitem[Koyama(2016)]{Koyama2016} Koyama, K.\ 2016, Reports on Progress in Physics, 79, 046902

\bibitem[Landy \& Szalay(1993)]{LSestimator} Landy, S.~D. \& Szalay, A.~S.\ 1993, \apj, 412, 64

\bibitem[Lewis(2019)]{Getdist} Lewis, A.\ 2019, arXiv:1910.13970

\bibitem[Lewis \& Bridle(2002)]{CosmoMC} Lewis, A. \& Bridle, S.\ 2002, \prd, 66, 103511

\bibitem[LSST Science Collaboration et al.(2009)]{LSST} LSST Science Collaboration, Abell, P.~A., Allison, J., et al.\ 2009, arXiv:0912.0201

\bibitem[Mar{\'\i}n et al.(2016)]{Marin2016} Mar{\'\i}n, F.~A., Beutler, F., Blake, C., et al.\ 2016, \mnras, 455, 4046

\bibitem[Matsubara(2008)]{Matsubara2008} Matsubara, T.\ 2008, \prd, 78, 083519

\bibitem[McDonald \& Seljak(2009)]{McDonald2009} McDonald, P. \& Seljak, U.\ 2009, \jcap, 2009, 007

\bibitem[Mueller et al.(2019)]{Mueller2019} Mueller, E.-M., Percival, W.~J., \& Ruggeri, R.\ 2019, \mnras, 485, 4160

\bibitem[Nikoloudakis et al.(2013)]{Nikoloudakis2013} Nikoloudakis, N., Shanks, T., \& Sawangwit, U.\ 2013, \mnras, 429, 2032

\bibitem[Padmanabhan et al.(2012)]{Padmanabhan2012} Padmanabhan, N., Xu, X., Eisenstein, D.~J., et al.\ 2012, \mnras, 427, 2132

\bibitem[Peacock \& Dodds(1994)]{Peacock1994} Peacock, J.~A. \& Dodds, S.~J.\ 1994, \mnras, 267, 1020

\bibitem[Peacock et al.(2001)]{Peacock2001} Peacock, J.~A., Cole, S., Norberg, P., et al.\ 2001, \nat, 410, 169

\bibitem[Reid \& White(2011)]{Reid2011} Reid, B.~A. \& White, M.\ 2011, \mnras, 417, 1913

\bibitem[Percival et al.(2001)]{Percival2001} Percival, W.~J., Baugh, C.~M., Bland-Hawthorn, J., et al.\ 2001, \mnras, 327, 1297

\bibitem[Percival et al.(2007)]{Percival2007} Percival, W.~J., Cole, S., Eisenstein, D.~J., et al.\ 2007, \mnras, 381, 1053


\bibitem[Perlmutter et al.(1999)]{Perlmutter1999} Perlmutter, S., Aldering, G., Goldhaber, G., et al.\ 1999, \apj, 517, 565

\bibitem[Planck Collaboration et al.(2018)]{PLC2018} Planck Collaboration, Aghanim, N., Akrami, Y., et al.\ 2018, arXiv:1807.06209

\bibitem[Riess et al.(1998)]{Riess1998} Riess, A.~G., Filippenko, A.~V., Challis, P., et al.\ 1998, \aj, 116, 1009

\bibitem[Riess et al.(2007)]{Riess2007} Riess, A.~G., Strolger, L.-G., Casertano, S., et al.\ 2007, \apj, 659, 98

\bibitem[Ross et al.(2013)]{Ross2013} Ross, A.~J., Percival, W.~J., Carnero, A., et al.\ 2013, \mnras, 428, 1116

\bibitem[Ross et al.(2014)]{Ross2014} Ross, A.~J., Samushia, L., Burden, A., et al.\ 2014, \mnras, 437, 1109

\bibitem[Ross et al.(2015)]{Ross2015} Ross, A.~J., Samushia, L., Howlett, C., et al.\ 2015, \mnras, 449, 835

\bibitem[Scoccimarro(2004)]{Scoccimarro2004} Scoccimarro, R.\ 2004, \prd, 70, 083007

\bibitem[Scoccimarro(2015)]{Scoccimarro2015} Scoccimarro, R.\ 2015, \prd, 92, 083532

\bibitem[Scolnic et al.(2018)]{Pantheon} Scolnic, D.~M., Jones, D.~O., Rest, A., et al.\ 2018, \apj, 859, 101

\bibitem[Seljak(2009)]{Seljak2009} Seljak, U.\ 2009, \prl, 102, 021302

\bibitem[Shectman et al.(1996)]{LCRS} Shectman, S.~A., Landy, S.~D., Oemler, A., et al.\ 1996, \apj, 470, 172

\bibitem[Slosar et al.(2008)]{Slosar2008} Slosar, A., Hirata, C., Seljak, U., et al.\ 2008, \jcap, 2008, 031



\bibitem[Sullivan et al.(2011)]{SNLS3} Sullivan, M., Guy, J., Conley, A., et al.\ 2011, \apj, 737, 102

\bibitem[Suzuki et al.(2012)]{Union} Suzuki, N., Rubin, D., Lidman, C., et al.\ 2012, \apj, 746, 85

\bibitem[Tamone et al.(2020)]{Tamone2020} Tamone, A., Raichoor, A., Zhao, C., et al.\ 2020, arXiv:2007.09009

\bibitem[Taruya et al.(2012)]{RegPT} Taruya, A., Bernardeau, F., Nishimichi, T., et al.\ 2012, \prd, 86, 103528

\bibitem[Taruya et al.(2010)]{Taruya2010} Taruya, A., Nishimichi, T., \& Saito, S.\ 2010, \prd, 82, 063522

\bibitem[Tegmark et al.(1997)]{Tegmark1997} Tegmark, M., Taylor, A.~N., \& Heavens, A.~F.\ 1997, \apj, 480, 22

\bibitem[Viljoen et al.(2020)]{Viljoen2020} Viljoen, J.-A., Fonseca, J., \& Maartens, R.\ 2020, arXiv:2007.04656

\bibitem[Wands(2010)]{Wands2010} Wands, D.\ 2010, Classical and Quantum Gravity, 27, 124002

\bibitem[Wang et al.(2014)]{Wang2014} Wang, L., Reid, B., \& White, M.\ 2014, \mnras, 437, 588

\bibitem[Wang et al.(2020)]{Wang2020} Wang, Y., Zhao, G.-B., Zhao, C., et al.\ 2020, \mnras, doi:10.1093/mnras/staa2593

\bibitem[White et al.(2009)]{White2009} White, M., Song, Y.-S., \& Percival, W.~J.\ 2009, \mnras, 397, 1348

\bibitem[Wilson et al.(2017)]{Wilson2017} Wilson, M.~J., Peacock, J.~A., Taylor, A.~N., et al.\ 2017, \mnras, 464, 3121

\bibitem[Yamamoto et al.(2006)]{Yamaestimator} Yamamoto, K., Nakamichi, M., Kamino, A., et al.\ 2006, \pasj, 58, 93

\bibitem[Yamauchi et al.(2014)]{Yamauchi2014} Yamauchi, D., Takahashi, K., \& Oguri, M.\ 2014, \prd, 90, 083520

\bibitem[York et al.(2000)]{sdss} York, D.~G., Adelman, J., Anderson, J.~E., et al.\ 2000, \aj, 120, 1579

\bibitem[Zhao et al.(2020)]{EZmock} Zhao, C., Chuang, C.-H., Bautista, J., et al.\ 2020, arXiv:2007.08997

\bibitem[Zhao et al.(2016)]{Zhao2016} Zhao, G.-B., Wang, Y., Ross, A.~J., et al.\ 2016, \mnras, 457, 2377

\bibitem[Zhao et al.(2019)]{DR14QGBZ} Zhao, G.-B., Wang, Y., Saito, S., et al.\ 2019, \mnras, 482, 3497

\bibitem[Zhao et al.(2020)]{Zhao2020} Zhao, G.-B., Wang, Y., Taruya, A., et al.\ 2020, arXiv:2007.09011

\bibitem[Zheng et al.(2019)]{Zheng2019} Zheng, J., Zhao, G.-B., Li, J., et al.\ 2019, \mnras, 484, 442


\end{thebibliography}
\end{document}